\documentclass{ws-ijmpd}

\input{tcilatex}

\begin{document}

\title{CHAOTIC MONOPOLE INTERACTIONS AND \\
VACUUM DISORDER}
\author{HILMAR FORKEL}
\address{Institut f\"{u}r Physik, Humboldt-Universit\"{a}t zu Berlin\\
D-12489 Berlin, Germany \\
forkel@ift.unesp.br}

\maketitle

\begin{abstract}
We study chaotic regions in the phase space of classical non-Abelian gauge
theory, focusing particularly on those which determine the low-energy
interactions between BPS\ monopoles, and comment on the relevance of the
obtained results for long-standing speculations which relate classical
Yang-Mills chaos to the disordered quantum vacuum and quark confinement.
\end{abstract}

\keywords{Gauge theory; chaos; magnetic monopoles.}

\markboth{Hilmar Forkel}{Chaotic Monopole Interactions}

%
\catchline{}{}{}{}{} %

\section{Why study chaos in (semi-) classical gauge theories?}

Deterministic chaos in the time evolution of classical non-Abelian (and
hence nonlinear) gauge theories has been investigated for about thirty years%
\cite{bir94}. This ongoing endeavour has several motivations. On the more
conceptual side, the chaotic behavior of gauge theories reveals typical
signatures of quantum chaos, visible e.g. in the distribution of
nearest-neighbor level spacings of lattice Dirac spectra according to
Gaussian matrix ensembles\cite{hal95}, shows a continuous cascading of the
dynamical degrees of freedom (and their energy) towards the ultraviolet
during time evolution\cite{mue92} and has potential implications for the
continuum limit of lattice gauge theories\cite{nie96}. One of the oldest and
perhaps most important fundamental motivations for investigating chaos in
non-Abelian gauge theories, furthermore, was to shed light on its potential
role in the QCD confinement mechanism\cite{bir94,bas79}. Much of the work
with more phenomenological focus, on the other hand, attempts to gain
insight into otherwise hardly accessible non-equilibrium processes by
exploiting the increasingly classical behavior of long-wavelength fields
with growing temperature. Such processes are of great interest in the
context of current experimental heavy-ion programs at RHIC and LHC which
create and analyze matter under extreme conditions. The so far best studied
examples are particle production from collective fields\cite{bir94} and
relations between the maximal Lyapunov exponents of the classical dynamics
and the damping or\ thermalization rates of hot gauge systems\cite%
{bir95,hei97}.

Astrophysical and cosmological applications, motivated by the classical
thermodynamics of hot and dense gauge theories as well, range from the
description of stellar interiors to the role of chaos during semi-classical
evolution phases of the early universe. Interesting examples are
topological\ structure formation and its impact on baryon number violating
production processes in the Standard Model at temperatures around the
electroweak phase transition\cite{pro97}. The chaoticity properties of
particle motion in various curved spacetimes have also been investigated
(see e.g. Ref. \refcite{vam96}).

A significant part of the results on gauge-theory chaos was obtained in
Yang-Mills-Higgs (YMH) theories. These are well suited for the study of
transitions from quantum to classical chaotic behavior since their
weak-coupling and semiclassical limits are controllable at all length scales
(in contrast to those of pure YM theories and QCD). The chaotic behavior of
the YMH theory with Higgs fields in the fundamental representation of the
gauge group (which forms a part of the electroweak sector of the standard
model) is e.g. of relevance for the mentioned baryon number violating
processes. YMH theories with Higgs fields in the adjoint representation, on
the other hand, appear in grand-unified theories and are of special interest
because they sustain stable, finite-action monopole solutions\cite{tho74}.

The latter will be a main focus of this article. More specifically, we will
study the regular and chaotic regimes in the low-energy dynamics of the
two-monopole system, and the transitions between them. This dynamics is
amenable to an enormous but controlled dimensional reduction of the relevant
phase space -- the geodesic approximation -- which we will outline in the
following section. In the subsequent Sec. \ref{chao} we quantify the
chaoticity of the monopole-monopole interactions in various phase space
regions, and in Sec. \ref{conf} we will address potential implications for
vacuum disorder and the quark confinement mechanism in \emph{quantum}
Yang-Mills theory and QCD. Section \ref{concl}, finally, contains a summary
and some conclusions.

\section{Geodesic two-monopole dynamics}

In the following two sections we are going to report on a recent
investigation of regular and\ chaotic low-energy interactions among
monopoles in the simplest possible setting\cite{far05}, i.e. between two
electrically charged magnetic Bogomol'nyi-Prasad-Sommerfield (BPS) monopoles%
\cite{pra75} or dyons\footnote{%
These solutions bear interesting similarites to the BPS dyon constituents of
caloron solutions with nontrivial holonomy\cite{lee98} which consist of a
BPS monopole-antimonopole pair (for $N_{c}=2$).} which solve the SU$\left(
2\right) $ YMH Bogomol'nyi equation\cite{pra75}%
\begin{equation}
B_{i}^{a}=\frac{1}{2}\varepsilon _{ijk}F_{jk}^{a}=\pm \left( \delta
^{ac}\partial _{i}+g\varepsilon ^{abc}A_{i}^{b}\right) \Phi ^{c}  \label{beq}
\end{equation}%
(where $F_{\mu \nu }^{a}$ is the field strength tensor of the gauge field $%
A_{\mu }^{a}$ and $\Phi ^{a}$ is the (adjoint) Higgs field).

The two-BPS monopole system is prototypical for physically interesting
subsystems of the gauge dynamics whose spatially varying fields -- here the
solitonic monopoles with topologically induced magnetic charge -- and time
evolution can be studied without invoking either uncontrolled approximations
(as e.g. the drastic ``homogeneous approximation'' used in the pioneering
studies\cite{bas79}) or requiring the elaborate lattice solution of the
full, hyperbolic Yang-Mills-Higgs equations\cite{bir94,mue92,bir95,hei97}.
This is because the low-energy time evolution of the dyon pair is accurately
described by the geodesic motion of a point on the manifold which the few
collective degrees of freedom of the two-monopole solution span\cite{NM:82},
i.e. it is governed by ordinary differential equations.

The basis for this geodesic approximation is that all two-monopole solutions
form a family whose members are parametrized by continuous collective
coordinates or ``moduli'' $x^{\alpha }$. (For the one-monopole solution,
e.g., these are the three position coordinates of the center and an overall
phase angle.) The corresponding moduli space $M_{2}$ is a manifold whose
metric is induced by the metric on the space of all finite-energy field
configurations, i.e. by the kinetic terms of the YMH Lagrangian, and known
explicitly. Owing to energy conservation and the degeneracy of all static
two-monopole solutions, the low-energy dynamics of two BPS dyons then
describes \emph{geodesic} motion of the associated point on $M_{2}$. After
separating the center-of-mass motion and an overall phase (whose time
dependence is associated with the total electric charge), the remaining
four-dimensional internal part $M_{2}^{\left( 0\right) }$ of the moduli
space can be parametrized by three Euler angles $\vartheta $, $\varphi $ and 
$\psi $, which determine the orientation of the two-monopole system, and the
distance variable $\varrho $ which measures (at large $\varrho $) the
separation between the two centers. The metric $g_{\alpha \beta }^{\left( 
\text{AH}\right) }$ on $M_{2}^{\left( 0\right) }$ has been constructed
explicitly by Atiyah and Hitchin\cite{AH:85}, and the resulting internal
Lagrangian is that of a non-rigid body with distance-dependent ``moments of
inertia'' around the body-fixed axes\cite{GM:86},%
\begin{equation}
L_{\text{AH}}\left( x,\dot{x}\right) =\frac{m}{2}g_{\alpha \beta }^{\left( 
\text{AH}\right) }\left( x\right) \dot{x}^{\alpha }\dot{x}^{\beta }=L_{\text{%
AH}}\left( \varrho ,\vartheta ,\psi ,\dot{\varrho},\dot{\vartheta},\dot{\psi}%
\right) ,  \label{geom}
\end{equation}%
where $m$ is the reduced mass of the monopoles. Physically, the validity of
the geodesic approximation implies that at small velocities (compared to the
velocity of light) internal excitations (vibrations) and deexcitations
(radiation) of the dyons can be neglected, i.e. the monopole pair adapts
adiabatically to its interactions by deforming reversibly and scattering
elastically.

\section{Regularity and chaos in monopole interactions \label{chao}}

The possibility for chaotic behavior depends on the number of integrals of
the motion of the underlying dynamics, which must be less than the number of
degrees of freedom. For the geodesic dynamics (\ref{geom}) of the
two-monopole system with its four degrees of freedom, three independently
conserved quantities are known explicitly. Those are the total angular
momentum $\mathcal{M}^{2}$, the energy $E_{\text{AH}}$\ and the generalized
momentum $p_{\varphi }$ conjugate to the coordinate $\varphi $ which is
cyclic, i.e. does not appear explicitly in the Lagrangian (\ref{geom}). For
the two-monopole dynamics to be (Liouville) integrable would therefore
require the existence of minimally one additional constant of the motion.
Such a fourth conserved quantity indeed exists at least if the two monopoles
remain infinitely separated, since then their \emph{individual} electric
charge is conserved. This situation changes when the two dyons begin to
approach each other and only their total charge remains conserved during
Higgs-induced charge exchange. Then chaotic motion becomes possible and
first numerical evidence for its existence in the two-dyon phase space was
gathered in Refs. \refcite{TR:88a}.

In Ref. \refcite{far05} we have then systematically analyzed regular and
chaotic two-monopole motions on the basis of thirteen long-time phase space
trajectories for which four-dimensional time series were generated by
numerically integrating the equations of motion with high accuracy over
typically $2^{25}$ time steps. The initial data sets were chosen to cover a
representative range of motion patterns and to explore the low-energy dyon
interactions at different strengths. The resulting set of orbits includes
sequences of trajectories whose decreasing minimal dyon separations
interpolate between asymptotic dyon distances, where charge exchange becomes
ineffective and the geodesic dynamics integrable, and relatively small
minimal separations for which the interactions are expected to become
non-integrable. Hence these orbits allow to map out the order-chaos
transition in the two-monopole system.

The first part of our analysis consisted in constructing the Poincar\'{e}
sections of this orbit set from the Hamiltonian on the reduced
four-dimensional phase space in which $p_{\varphi }$ and $\mathcal{M}^{2}$
are conserved and act as fixed ``external'' parameters. Energy conservation
then constrains all orbits to three-dimensional hypersurfaces. (Numerically, 
$E$, $\mathcal{M}^{2}$ and $p_{\varphi }$ were conserved up to deviations of
order $10^{-12}$.) Orbits with weak initial Coulomb attraction between the
dyons cover a rather large range of $\varrho $ values with relatively
moderate variations of the radial velocities which stay well inside the
asymptotic region of approximately (or KAM-)\ integrable motion. Hence their
Poincar\'{e} sections (in the $\left( \varrho ,p_{\varrho }\right) $ plane
at $\mathcal{M}_{1}=0$) consist of one-dimensional, continuous closed curves
corresponding to quasiperiodic motions. When increasing the initial Coulomb
attraction, the variations in dyon distance become smaller (tighter orbits)
while their relative momenta vary more strongly. The increased attraction
also brings the dyons closer together. From a certain minimal inter-dyon
distance $\varrho _{\min }\sim 2\pi $ onward the section visibly spreads out
into a broadly distributed scatter of points which eventually fills the $%
\left( \varrho ,p_{\varrho }\right) $ plane. This is a typical signature for
the corresponding aperiodic orbit to have become chaotic.

In order to investigate this chaoticity further, we have calculated
high-resolution power spectra of the momentum conjugate to the dyon
separation for selected orbits. This allows for a more accurate distinction
between quasiperiodic and chaotic motion patterns and yields quantitative
information about the underlying scales. The resulting spectra indeed
clearly separate quasiperiodic from irregular behavior. Two of the orbits,
as expected those with the maximal initial Coulomb force between the dyons,
were identified as chaotic. This substantially increases previous evidence
that the relative low-energy motion of two BPS dyons admits, apart from the
asymptotic $\varrho \rightarrow \infty $ region, only three independent
conserved quantities and turns out to be genuinely non-integrable.

In addition, the power spectra characterize quasiperiodic dyon-pair orbits
(i.e. those which remain close enough to the asymptotic region)
quantitatively by establishing the number of fundamental modes (two),
determining their frequencies and yielding the strength distribution over
their various harmonics\cite{far05}. The restriction to the minimal number
of quasiperiodic modes is rather widespread among nonlinear dynamical
systems if they are sufficiently strongly coupled. The common expectation
that nonlinear couplings between more than two fundamental modes
increasingly turn quasiperiodicity into chaos may therefore apply to the
two-dyon system as well and explain why we have encountered only
two-mode-quasiperiodic and chaotic trajectories.

In contrast to their rather complete specification of quasiperiodic motion
patterns, power spectra do extract relatively little pertinent quantitative
information from aperiodic orbits. Hence we have additionally calculated
those quantities which perhaps most directly quantify the chaoticity of
irregular motion patterns, i.e. the maximal Lyapunov exponents, for a
suitable set of orbits. As expected, the Lyapunov exponents of orbits
previously identified as quasiperiodic were found to vanish. The two orbits
with an irregular broadband power spectrum, on the other hand, turned out to
have finite and positive maximal Lyapunov exponents whose values were
approximately determined as $\mathsf{L}_{\text{max,1}}\sim 0.02$ and $%
\mathsf{L}_{\text{max,2}}\sim 0.008$. These exponents provide our most
unequivocal and quantitative evidence for the chaoticity of the dyon-dyon
interactions. The orbit with the smaller Lyapunov exponent, furthermore,
shows signs of intermittent behavior\cite{far05}.

\section{Chaotic monopole systems and confinement}

\label{conf}

In the following section we will summarize several ideas and speculations on
the potential relevance of the discovered two-monopole chaoticity for vacuum
disorder and confinement\cite{mil,gre07} in quantum YM and YMH theories.

We start by recalling the long-standing conjecture that the vacuum of
non-Abelian gauge theories, when undergoing a transition from weakly to
strongly coupled fields, also undergoes an order-disorder transition, and
that the strongly coupled QCD vacuum is populated by highly irregular color
field configurations\cite{bir94}. In the limit of a large number of colors,
in particular, a vacuum made of random Yang-Mills fields is known to be a
necessary and sufficient condition for quark confinement\cite{ole82}. From
the outset, one of the motivations for investigating chaos in non-Abelian
gauge theories was therefore to shed light on its potential role in the
confinement mechanism\cite{bir94,bas79}. More recently, a numerical
investigation of the classical time evolution of Yang-Mills field
configurations generated by finite-temperature (quantum) lattice simulations%
\cite{bir97} has provided evidence for the confining strong-coupling phase
to be indeed substantially more chaotic than its weakly coupled counterpart:
as a function of increasing coupling the maximal Lyapunov exponent undergoes
a sharp transition to larger values around the confinement transition.

Moreover, the instability of constant color-magnetic vacuum fields\cite%
{sav77} made it natural to assume that both gauge invariance and stability
of the physical vacuum may be restored by disordering color-magnetic
background fields. Under the gluonic infrared degrees of freedom\cite{for07}
envisioned to exhibit (and maybe cause) this disorder are random domains as
well as vacuum populations of quasi-randomly distributed, percolating center
vortices or monopoles\cite{gre07}.

The pivotal role which non-Abelian monopoles and their chaotic interactions
may play in the context of quark confinement in QCD is further presaged in $%
\mathcal{N}=2$ supersymmetric Yang-Mills theory in 3+1 dimensions\cite{sei94}
where BPS monopoles indeed realize a non-Abelian version of the classic 't
Hooft-Mandelstam dual superconductor confinement mechanism\cite{tho76}. In
this scenario the condensation of magnetic BPS monopole charges screens
color\ magnetic charges and confines color electric charges by the dual
Mei\ss ner effect. Similar scenarios, in which the condensation of
monopole-like objects plays a key role, are expected to unfold in more
physical gauge theories as well. As a case in point, in the 2+1 dimensional
Yang-Mill-Higgs model 't Hooft-Polyakov monopoles (a generalization of BPS\
monopoles which play the role of instantons in this case) generate ``weak
confinement'' by forming a\ monopole antimonopole plasma, as shown by
Polyakov\cite{pol77}. In 3+1 dimensional Yang-Mills theories, furthermore,
there is lattice evidence for the condensation of Abelian-projected
monopoles to generate the bulk of the string tension\cite{che97}. (According
to an interesting recent suggestion, the ``active'' monopoles might actually
be BPS dyon constituents of caloron solutions with nontrivial holonomy\cite%
{dia08}.)

In light of the above arguments it is tempting to speculate that a
disordered ensemble of monopoles (and anti-monopoles)\ in a semiclassical
vacuum may be generated by the classically chaotic low-energy interactions
among individual monopole pairs which we have investigated above. Below we
will suggest two approaches towards pursuing and testing such ideas in a
more quantitative fashion.

\section{Summary and conclusions \label{concl}}

We have discussed chaotic regions in the classical phase space of
non-Abelian gauge theories and studied, in particular, regular and chaotic
motion patterns of two interacting BPS monopoles at low energies. Our
analysis is based on a representative set of long-time phase-space
trajectories in the geodesic approximation and intended to survey the
order-chaos transition and to characterize the quasiperiodic and chaotic
behavior quantitatively.

The observed changes in the dimension of the trajectories' Poincar\'{e}
sections (from one to two) provide clear evidence for transitions from
quasiperiodic to chaotic motion when the monopoles come close enough to each
other. These results were confirmed and quantified by the analysis of
high-resolution power spectra and Lyapunov exponents for selected orbits.
The\ obtained values for the maximal Lyapunov exponents contain information
on the relaxation time and thermalization (damping)\ rate of a
non-equilibrium dyon system at sufficiently high temperatures. Taken
together, our results provide convincing evidence for and a quantitative
description of both quasiperiodic and chaotic regions in the low-energy
phase space of two BPS dyons. They also show that no more than the three
explicitly known integrals of the motion are conserved by the geodesic
forces between the dyons.

Since the motion of free\ dyons is integrable, the chaotic behavior analyzed
above can be uniquely traced to the interactions between the dyons. This
raises hopes that our quantitative understanding of the chaotic dyon-dyon
interactions may also generate new insights into the disorder of interacting
monopole ensembles of the type which are expected to populate the vacuum of
the strong interactions. For sufficiently dilute systems, expansions in the
monopole density and more sophisticated many-body techniques may e.g.
provide a quantitative treatment of chaotic multi-monopole systems. Another
option would be the technically challenging extension of the geodesic
approximation to approximate multi-monopole-antimonopole solutions with
their more complex interactions.

\section*{Acknowledgments}

The author would like to thank Ricardo Fariello and Gast\~{a}o Krein for
their collaboration on this project and the funding agencies Funda\c{c}\~{a}%
o de Amparo \`{a} Pesquisa do Estado de S\~{a}o Paulo (FAPESP), Conselho
Nacional de Desenvolvimento Cient\'{\i}fico e Tecnol\'{o}gico (CNPq) and
Deutsche Forschungsgemeinschaft (DFG) for finanical support.


\end{document}